# Health Information Systems (HIS): Concept and Technology

Mohd. Nabil Almunawar
Universiti Brunei Darussalam
nabil.almunawar@ubd.edu.bn

Muhammad Anshari
Universiti Brunei Darussalam
anshari@yahoo.com

*Abstract*— A health information system (HIS) is the intersection of between healthcare's business process, and information systems to deliver better healthcare services. The nature of healthcare industry, which is highly influenced by economic, social, politic, and technological factors, has changed over time. This paper will address some important concepts of healthcare and related terminologies to provide a holistic view for HIS. Related technological milestones and major events are briefly summarized. The trends and rapid development of health information technologies are also discussed.

*Keywords; Health Information Systems (HIS), Information Communication Technology (ICT), Healthcare Organization, Web 2.0*

## I. INTRODUCTION

Nowadays, the widespread use of Information and Communication Technologies (ICT) has permeated almost all aspects of life including the healthcare sector. HIS was introduced to fully utilize especially the Internet in providing better healthcare. Health information systems are frequently refers to the interaction between people, process and technology to support operations, management in delivering essential information in order to improve the quality of healthcare services. Similar to any other industries, the nature of healthcare industry has changed over time from a relatively stable industry to a dynamic one. And health information systems have evolved through several different technologies.

Haux (2006) describes systems that process data and provides information and knowledge in healthcare environments as health information systems. Hospital information systems are just an instance of health information systems, in which a hospital is the healthcare environment as well as healthcare institution. The aim of health information systems is to contribute to a high-quality, efficient patient care.

Some of terminologies related to HIS are as follows. Health Informatics is the field that concerns itself with the cognitive, information processing, and communication tasks of medical practice, education, and research including the information science and technology that supports those tasks. Health informatics tools include computers as well as clinical guidelines, formal medical terminologies, and information and communication systems. In other words, it emphasis is on clinical and biomedical applications with added possibility of the integrating clinical components either among themselves or to more administrative-type health information systems (Conrick, 2006). Additionally, Health information technology is the application of information processing involving both computer hardware and software that deals with the storage, retrieval, sharing, and use of health care information, data, and knowledge for communication and decision making (Goldschmidt, 2005).

Another important terminology in HIS is Electronic Medical Records (EMR), it resides at the centre of any health information systems. EMR is a medical record in a digital format, whereas electronic health record (EHR) refers to an individual patient's medical record in a digital format. EHR systems coordinate the storage and retrieval of individual records with the aid of computers, which are usually accessed on a computer, often through a computer network. One of the important trends is the move towards a universal electronic patient record (EPR). EPR is defined as electronically stored health information about one individual uniquely identified by an identifier. Essentially EPR technology entails capturing, storing, retrieving, transmitting, and manipulating patient-specific, healthcare related data singly and comprehensively, including clinical, administrative, and biographical data (Protti et al., 2009).

One of the most interesting aspects of HIS is how to manage the relationship between healthcare providers and patients. Fostering good relationship with customer (patient) will retain them and attract them to become loyal customers, create greater mutual understanding, trust, and satisfaction. In addition, a good relationship will encourage patient's involvement in decision making (Richard and Ronald, 2008). A good relation will foster effective communication which is often associated with improved physical health, more effective chronic disease management, and better health related quality of life (Arora, 2003). Managing relationship must continually develop and grow. A good relationship is a dynamic one that the organization become alert to and aware of changing needs. In turn, when healthcare organizations manage well, patients will want to come back, because loyalty and trust are built. They will know that if they present a difficulty, the organizations will resolve professionally.

The rest of the paper is organized as follows. In sections 2, we discuss history of health information systems. In section 3, we focus on future technological trends of HIS and finally section 4 is the conclusion.

## II. HISTORY

The use of ICT in healthcare is not new. Deploying ICT in healthcare environment has helped healthcare professionals to improve the efficiency and effectiveness of healthcare services. Healthcare information systems that can record and locate important information quickly have become a standard practice in many healthcare organizations. While, Haux (2006) summarized the milestone of development for HIS were considered as important: (1) the shift from paper-based to computer-based processing and storage, as well as the increase of data in health care settings; (2) the shift from institution-centered departmental and, later, hospital information systems towards regional and global HIS; (3) the inclusion of patients and health consumers as HIS users, besides health care professionals and administrators; (4) the use of HIS data not only for patient care and administrative purposes, but also for health care planning as well as clinical and epidemiological research; (5) the shift from focusing mainly on technical HIS problems to those of change management as well as of strategic information management; (6) the shift from mainly alpha-numeric data in HIS to images and now also to data on the molecular level; (7) the steady increase of new technologies to be included, now starting to include ubiquitous computing environments and sensor-based technologies for health monitoring.

Healthcare is undergoing a paradigm shift, moving from 'Industrial Age Medicine to Information Age Healthcare' (Smith, 1997). This 'paradigm shift' is shaping healthcare systems (Haux et.al, 2002) and transforming the healthcare-patient relationship (Ball, 2001). For example, the World Wide Web has changed the way the public engage with health information (Powell et al., 2003).

TABLE I. Healthcare Information Systems by Industry Phase at USA (Bourke M.K, 1994)

| Industry phase | Data | Technology |
|---|---|---|
| 1945 – 1965 Govt. sponsored growth | Manual | Almost no Information Technology |
| 1965 – 1973 Medicare introduced | GL and AP (Finance) | Mainframes, Stand alone, No standard |
| 1973 – 1983 Disenchantment on many fronts | Utilization data, Profitability reporting | Minicomputers, PC, DBMS on Mainframe |
| 1983 – 1991 Diagnose Related Group introduced | Data collection dictated by external organization | PC networks, PC database |
| 1991 – 2000 Prospect of National health care | Product line, Market segment, Demographic segment | PC networks and database, AI, Data exchange |

The introduction of the personal computer (PC) during 1980s to 1990s, which later supported by Local Area Network (LAN) has made the use of health information systems widespread. . Health information systems providers began to reengineer their products and collaborated with other vendors to make it more open. For instant, IBM and Bexter entered joint venture, which subsumed the former Omega and Delta product (Bourke M.K, 1994). Database management systems equipped with query languages enabled other vendors to access database. Query language mediated interoperability and heterogeneous databases across vendors.

From the late 1980s, technology was developed in order to make the delivery of more tailored services and products at lower prices possible. During this period, healthcare organizations were shifting towards an integrated care. A trend towards open systems and object technologies has already been emerging during the '90s, institutional mergers and networks have made new concepts mandatory (Kuhn and Giuse, 2001). In US, the merger of hospitals and individual practices into large integrated healthcare networks has been described as a dominant trend (Teich, 1998), while in Europe has been described as a decentralized network of healthcare delivery institutions that slowly replaces hospitals as centers of care delivery (Iakovidis , 1998). The trend towards clinical computing and a patient centered computer-based record could be seen worldwide in 1980s (Ball, 1999). As predicted, the hospital information system of earlier decades which was mainly administrative functionality had become much more focused on the clinical perspective and the patient record, while becoming more open in a technological as well as an organizational sense. It is now understood that data, not systems, is what counts (Tuttle, 1999). Moreover, the critical issue is people – not technology, and technology is the enabler, not the driver (Ball, 1999).

Advancement in the Internet technology in 1990s facilitated to process, store, retrieve and disseminate data and information remotely anytime and anywhere. It was dramatically changing the ways in which the healthcare organization operates. The Internet has enabled customer to access information on the products and services they are considering from other users, not only from producer. Customer demands changed because they have been empowered by the amounts of information available to them.

Starting from 1991, the development of IT has marked the new generation of networked technology which affected all areas including healthcare service and technology. It could be influenced by the increase in the connectivity of networks and the ability of database management systems to execute multitasks across multiple networks and databases. The broad use of technologies including expert systems, voice recognition, electronic imaging, and voice synthesis were introduced during in 1990s. Networking technologies and database management systems were able to incorporate better healthcare service, faster in response, and easier to meet increasing demands for the system integration.

The power of the Internet for both customers and businesses became visible in the mid-1990s when both e-commerce and online review sites became increasingly popular ways to do business and communicate. It made information easily available directly from the providers via their websites. The Internet technology such as the Web also gave a way for customers to share their opinions on the products and services that they buy which are often more conversant than the information on the products and services provided by ther producers. This triggered a significant change on how companies treated customers. As a result, companies has started engage customer listened to their concerns and involved them in improving the quality products and services.

The Web has become a popular channel to deliver information products or services, including those in healthcare industry. The use of web browsers for clinical workstations has increase significantly, as browsers offer a simple and intuitive user interface (Kuhn and Giuse, 2001). The feasibility of building web interfaces to clinical information systems has been shown in 1996 (Sittig, 1996), and reports of successful projects followed. WebCIS is an example of implementing a web server atop clinical information system architecture with a central data repository (Hripcsak, 1999). The W3-EMRS project explored web technology as a framework for integrating heterogeneous information systems, and a web-based virtual repository approach was successfully implemented (Wang et.al, 1997). The Web, with its vast array of health information resources, is a desirable means that holds great promise for reaching customer (patient) anytime anywhere. Web based knowledge may fill information needs, though the quality and relevance of resources vary widely.

### III. FUTURE TRENDS

Today, every healthcare organization depends on ICT in every level of activities. Nowadays, the healthcare relies on process application and information streamline to create value for every facet of its delivery. The aim of this paper is to briefly summarize on the past and current health information systems and to identify few emerging trends and research in HIS. The foremost observations to be drawn from previous sections include; concepts and terminologies related to health information systems' field, history of several generations of HIS, and concluded with the recent trend and development of tools and technologies in creating and managing HIS.

The vision of a paperless hospital is delineated as the embodiment of the future health information systems with the hope is that brings an improvement. promise of to be more reliable effective and efficient. The current status of HIS varies among countries. There are 193 countries that are a member of World Health Organization (WHO) in 2009; 114 of them participated in the global survey on e-health (WHO, 2011). Most developed countries have fully utilized HIS in their systems as they have the resources, expertise, and capital to implement them. While developing countries HIS have not been fully utilized yet.

As a business entity, a healthcare provider needs to provide the same standards of customer service as other business entities. The high expectation of customer service provided by healthcare organizations in the information age poses a serious challenge for healthcare providers as they have to make an exceptional impression on every customer. In the competitive commercial healthcare environment, negative experiences, and poor service leads customers to switch healthcare providers because poor service indicates inefficiency, higher cost and lower quality of care. No doubt, the adoption of HIS is believed to boost effectiveness and efficiency customer service in healthcare organizations.

The use of ICT in healthcare organizations has grown in the same pattern as compare to the larger industry landscape. The use of web technology, database management systems, and network infrastructure are part of ICT initiative that will affect of healthcare practice and administration.

One such trend is the slowly adoption of e-health systems toward the use of EMR. The systems move patient information from paper to electronic file formats so they can be easily and effectively managed. However, an interesting fact to be noted that the tendency of people to know more and actively participate in the health promotion, prevention, and care together with the rights that will become a standard legislature guide the development of information systems that support these tendencies. Thus, the trend is towards more involvement of patients or citizens in receiving information, in decision making and in responsibility for own health. The prime feature of this trend is the shifting from healthcare-institution centered care to the citizen-centered care emphasising on continuity of care from prevention to rehabilitation. This vision can be achieved through shared care which builds on health telematics networks and services, linking hospitals, laboratories, pharmacies, primary care and social centers offering to individuals a 'virtual healthcare centre' with a single point of entry. Furthermore, this vision implies provision of health services to homes with innovative services such as personal health monitoring and support systems and user-friendly information systems for supporting health education and awareness (Ilias, 1998).

Several examples of countries that have implemented HIS are Canada, Singapore and Australia to name a few. Canada has established e-health Ontario on March 2009 with three targeted strategies to improve; diabetes management, medication management and wait times. One of the examples of the service offered is ePrescribing under the medication management. It authorizes and transmits prescriptions from physicians and other prescribers to pharmacists and other dispensers (e-Health Ontario, 2009). It prevents medication error due to illegible prescribing and reduces fraud prescriptions. "Participating prescribers and pharmacies at both sites will continue electronically prescribing until a provincial Medication Management System is in place" (e-Health Ontario, 2009).

There are various emerging tools and technologies in creating and managing HIS. Semantic Web is an extension of the World Wide Web, offers a united approach to knowledge management and information processing by using standards to represent machine-interpretable information. Semantic Web technology helps computers and people to work better together by giving the contents well-defined meanings. The semantic Web has also drawn attention in the medical research communities (Cheung & Stephens, 2009). Semantic web services can support a service description language that can be used to enable an intelligent agent to behave more like a human user in locating suitable Web services. While, Web services are software components or applications, which interact using open XML and Internet technologies. These technologies are used for expressing application logic and information, and for transporting information as messages (Turner et al., 2004). They have significantly increased interest in Service oriented architectures (SOAs) (Erl, 2005). The benefits of Web services include loose coupling, ease of integration and ease of accessibility.

There are a number of health information support systems demonstrating applications of computer networks that tap into the vast array of health information available on the web. These systems are designed for patients with a health crisis or medical concern and for primary care providers. Patient-centered health information systems include models that use a variety of strategies to filter health information from the web to provide focused and tailored health information for concerns such as HIV/AIDS, asthma, smoking cessation, living with alcohol abuse, or stress (Brennan, Caldwell, Moore, Sreenath, & Jones, 1998; Gustafson et al., 1999). Physician-centered health information support systems use a variety of strategies to provide health information filtered from the WWW for physicians and patients/families on topics such as child health, safety, and the management of insulin-dependent diabetes (Porcelli & Lobach, 1999; Riva, Bellazzi, & Stefanelli, 1997).

A HealthGrid (HealthGrid, 2009) allows the gathering and sharing of many medical, health and clinical records/databanks maintained by disparate hospitals, health organizations, and drug companies. In other words, HealthGrid is an environment in which data of medical interest can be stored and made easily available to different actors in the healthcare system, physicians, allied professions, healthcare centres, administrators and, of course, patients and citizens in general. Also, the driving forces for the individual and commercial adoption of the VoIP are the significant cost savings, portability, and functionality that can be realized by switching some or all of their voice services to VoIP. Chen et al showed the integration of mobile health information system with VoIP technology in a wireless hospital (Cheng et al., 2008).

Recent development in ubiquitous computing is a paradigm shift since technology becomes virtually invisible in our lives. The ubiquitous computing environment will make possible new forms of organizing, communicating, working and living. However, ubiquitous computing systems create new risks to security and privacy. To organize the u-healthcare infrastructure, it is necessary to establish a context-aware framework appropriate for the wearable computer or small-sized portable personal computer in ubiquitous environment (Ko et al., 2007). The mobile health (m-health) a form of ubiquitous computing can be defined as mobile communications network technologies for healthcare (Istepanian et al., 2006). This concept represents the evolution of "traditional" e-health systems from desktop platforms and wired connections to the use of more compact devices and wireless connections in e-health systems.

The author concerns with the recent development of Web technology called Web 2.0 within e-health services to improve patients' satisfaction and their health literacy. Recently, the Web 2.0 tools such as *Facebook*, *Twitter*, *Myspace*, *Friendster*, *LinkedIn*, etc. have grown rapidly facilitating peer-to-peer collaboration, ease of participation, and ease of networking. However, the effects of Web 2.0, particularly in addressing the issues of healthcare services are still not much discussed. The main advantages of Web 2.0 are the linkage among people, ideas, processes, systems, contents and other organizational activities (Askool and Nakata, 2010). Therefore, Web 2.0 could affect healthcare business process like relationship between patients and healthcare providers as it is about engaging relationships, sharing experience & information, and collaboration.

The use of Web 2.0 in HIS system is equivalent to bringing patient expectation aligned with fashion of ICT in actual healthcare services. It offers new outlook either from patient or healthcare organization, and how they structure inter-relation between three distinct domains of objects; customer's expectation, advancement of ICT, and healthcare services. Each domain has unique features and characteristics which failing to respond appropriately may affect to business survivability and customer dissatisfaction.

With the advancement of ICT, Web 2.0 technology has brought a possibility to extend the service of HIS by enabling patients, patient's families, and community at large to participate more actively in the process of health promotion and education through social networking process.

We proposes integrated health information systems which include holistic approach of personal habit, physical activities, spiritual and emotional activities, and social support and social networks to be part of the systems. It is significant to improve customer satisfaction and health literacy in healthcare service to accommodate components and features of social networking capabilities, empowering patients, and availability of online health educator.

We propose our conceptual model for Social CRM in healthcare organizations. The model operates in the area of healthcare organization–patient relationships inclusive with social networks interaction, and how they possibly shared information to achieve health outcomes. It offers a starting point for identifying possible theoretical mechanisms that might account for ways in which Social CRM provides one-stop service for building relationship between healthcare organization, patients, and community at large.

The framework is developed from Enterprise Social Networks, Internal Social Networks, Listening tool interfaces, Social CRM systems within healthcare provider, and healthcare value configuration (value chain and value shop). Social Networks refers to the Web 2.0 technology that patient or his/her families may join any of them. It differentiates two social networks linkages to the patient or his/her family; they are Enterprises Social Networks and Internal Social Networks. The Enterprises Social Networks refers to external and popular Web 2.0 applications such as *Facebook*, *Twitter*, *LinkedIn*, *MySpace*, *Friendster*, etc which patient may belong to any of those for interaction. The dashed line connected enterprises social networks and CRM systems mean that none of those networks have control over the others directly, but constructive conversation and information from enterprises social networks should be captured for creating strategy, innovation, better service and at the same time responds accurately. Further, it proposes Internal Social Networks that operated, managed, and maintained within healthcare's infrastructure.

This is more targeted to internal patients/families within the healthcare to have conversation between patients/family within the same interest or health problem/ illness. For example, patient with diabetic would motivate to share his/her experiences, learning, and knowledge with other diabetic patients. Since patient/family who generates the contents of the Web, it can promote useful learning center for others, not only promoting health among each others, but also it could be the best place supporting group and sharing their experiences related to all issues such as; how the healthcare does a treatment, how much it will cost them, what insurance accepted by healthcare, how is the food and nutrition provided, etc. Therefore, this is generic group that will grow depends on the need of patients in that healthcare (Anshari and Nabil, 2011).

REFERENCES

[1] Anshari M, and Nabil M.A., (2011), "Evaluating CRM Implementation in Healthcare Organization", 2011 International Conference on Economics and Business Information, IPEDR vol.9 (2011) © (2011) IACSIT Press, Bangkok, Thailand

[2] Askool, S. S., and Nakata, K., 2010, Scoping Study to Identify Factors Influencing the Acceptance of Social CRM, Proceedings of the 2010 IEEE ICMIT P. 1055-1060

[3] Ball MJ, Peterson H, Douglas JV.The computerized patient record: a global view. MD Comput 1999; 16: 40-6.

[4] Ball, M.J., and Lillis, J., 2001. E-health: transforming the physician/patient relationship, Int. J. Med. Inform. 61 (1) 1–10.

[5] Bourke M.K, Strategy and Architecture of Health Care Information Systems, Springer-Verlag New York, 1994

[6] Brennan PF (1995) Characterizing the use of health care services delivered via computer networks. Journal of the American Medical Informatics Association 2:160–168.

[7] Cheng, P.H., Chen, S.J., Lai, J.S., & Lai, F. (2008). A collaborative knowledge management process for implementing healthcare enterprise information systems. IEICE Transactions on Information and Systems, E91-D(6), 1664-1672.

[8] Cheung, K.H., & Stephens, S. (2009). Semantic Web for health care and life sciences: A review of the state of the art. Briefings in Bioinformatics, 10(2), 111-113.

[9] Conrick, M. (2006). Health informatics: Transforming healthcare with technology. Thomson Social Science Press.

[10] E-Health Ontario, 2009, ePrescribing, viewed 1 March 2011, http://www.ehealthontario.on.ca/programs/ePrescribing.asp.

[11] Erl, T. (2005). Service-oriented architecture: Concepts, technology, and design. Prentice Hall.

[12] Goldschmidt, P.G. (2005). HIT and MIS: Implications of health information technology and medical information systems. Communications of the ACM, 48(10), 69-74.

[13] Gustafson D, Wise M, McTavish F, Taylor JO, Wolberg W, Stewart J, Smalley RV, Bosworth K (1993) Developing and pilot evaluation of a computer-based support system for women with breast cancer. Journal of Psychosocial Oncology 11(4):69–93

[14] Haux Reinhold, 2006, Health information systems – past, present, future, International Journal of Medical Informatics (2006) 75, 268—281.

[15] Haux, R., E. Ammenwerth, W. Herzog, P. Knaup, 2002, Health care in the information society. A prognosis for the year 2013, Int. J. Med. Inform. 66 (1–3) 3–21.

[16] HealthGrid Initiative (2009). Retrieved June 2009, from http://www.healthgrid.org

[17] Hripcsak G, Cimino JJ, Sengupta S.WebCIS: large scale deployment of a Web-based clinical information system. In: Lorenzi NM (ed). Proc AMIA Symp 1999; 804-8.

[18] Ilias Iakovidis, Towards personal health record: current situation, obstacles and trends in implementation of electronic healthcare record in Europe, International Journal of Medical Informatics, Volume 52, Issues 1-3, 1 October 1998, Pages 105-115, ISSN 1386-5056, DOI: 10.1016/S1386-5056(98)00129-4.

[19] Iakovidis I. Towards personal health record: current situation, obstacles and trends in implementation of electronic healthcare record in Europe. Int J Med Inf 1998; 52:105-15.

[20] Istepanian, R.H., Laxminarayan, S., & Pattichis, C.S. (Eds.) (2006). M-Health: Emerging mobile health systems. Springer.

[21] Joel J. P. C. Rodrigues and Binod Vaidya, Health Information Systems:Concepts, Methodologies, Tools and Applications in Health Information Systems: Concepts, Methodologies,Tools, and ApplicationsMedical Information Science Reference (an imprint of IGI Global), 2010

[22] Ko, E.J., Lee, H.J., & Lee, J.W. (2007). Ontology-based context modeling and reasoning for u-healthcare, IEICE Transactions on Information and Systems 2007 E90-D(8), 1262-1270.

[23] Kuhn, K.A., and Giuse D.A., 2001, From Hospital Information systems to Health Information Systems; Problems, Challanges, Perspectives, Method Inform Med 4/2001.

[24] Porcelli PJ, Lobach DF (1999) Proceedings American Medical Informatics Association 99 Annual Symposium, Integration of clinical decision support with on-line encounter documentation for well child care at the point of care (Hanley & Belfus, Washington, DC), pp 599–603

[25] Powell, J.A., M. Darvell, J.A. Gray,2003. The doctor, the patient and the world-wide web: how the internet is changing health care, J. R. Soc. Med. 96 (2) 74–76.

[26] Richard, L.S., and Ronald M.E. 2008. Lessons from Theory & Research on Clinician-Patient Communication In: Karen G., Barbara K.R, K.Viswanth (eds.) "Health Behavior and Health Education; Theory, Research, and Practice" 4th edition, (11) p 236-269 JOSSEY-BASS

[27] Riva A, Bellazzi R, Stefanelli M (1997) A Web-based system for intelligent management of diabetic patients. M.D. Computing 14:360–364

[28] Sittig DF, Kuperman GJ, Teich JM. WWWbased interfaces to clinical information systems: the state of the art. In: Cimino JJ (ed). Proc AMIA Annu Fall Symp 1996; 694-8.

[29] Smith, R.,1997. The future of healthcare systems, BMJ 314 (1997) 1495–1496.

[30] Teich JM. Clinical information systems for integrated healthcare networks. In: Chute CG (ed). Proc AMIA Symp 1998; 19-28.

[31] Turner, M., Zhu, F., Kotsiopoulos, I., Russel, M., Budgen, D., Bennet, K., Brereton, P., Keane, J., Layzell, P., & Rigby, M. (2004). Using web service technologies to create an information broker: an experience report. In Proceedings of the 26th International Conference on Software Engineering (ICSE'04) (pp.552-561). IEEE.

[32] Tuttle MS. Information technology outside health care: what does it matter to us? J Am Med Inform Assoc 1999; 6: 354-60.

[33] Wang K, van Wingerde FJ, Bradshaw K,Szolovits P, Kohane I. A Java-based multiinstitutional medical information retrieval system. In: Masys DR (ed) Proc AMIA Annu Fall Symp 1997; 538-42.

[34] World Health Organization (2011).Publications. Retrieved March 4, 2011, from http://www.who.int/goe/publications/en/